\documentstyle [12pt]{article} 
\begin{document}
\newcommand{\vm}{\vspace{0.2cm}}
\newcommand{\vl}{\vspace{0.4cm}}

\title{p-Adic description of Higgs mechanism I: p-Adic square root and p-adic
light cone }
\author{Matti Pitk\"anen\\
Torkkelinkatu 21 B 39, 00530, Helsinki, FINLAND}
\date{6.10. 1994}
\maketitle

\newpage

\begin{center}
Abstract
\end{center}

\vl

This paper is the first one in the series devoted to
 the calculation of particle mass spectrum in Topological
GeometroDynamics. In this paper p-adic conformal field theory limit of TGD
is formulated.  TGD Universe is critical at quantum level and the idea is
to realize criticality via  conformal invariance.  Ordinary real numbers
do not allow this but if one assumes that in long length scales p-adic
topology replaces real topology as effective topology situation changes.
The existence of square root in the vicinity of p-adic real axis  implies
4-dimensional algebraic  extension of p-adic numbers ($p>2$), which can be
regarded as p-adic counterpart  of light cone and consists of
convergence-cubes of p-adic square root function. Later work has demonstrated
that convergence cubes of square root function serve as natural quantization
volumes    in p-adic field theory limit of TGD.

\newpage
\tableofcontents

\newpage

\section{Introduction}

Quantum TGD has developed during past sixteen years from  an attractive
idea for realizing Poincare invariant gravitation to a quantitative theory
of basic interactions.  The basic features of TGD  \cite{TGD,padTGD}
relevant for this work are following. \\ a) TGD spacetime is surface of
space $H=M^{4}_+\times CP_2$.   The topology of 2-space is generalized to
the concept of topological condensate: particle like 3-surfaces are glued
by topological sum operation  to larger 3-surface, which in turned are
glued to larger surface and a many  sheeted fractal like structure
results. 3-surfaces can have boundaries. Even the boundaries of
macroscopic objects correspond to boundaries of 3-surface.   3-space need
not be even connected:  topological evaporation or the formation  of Baby
Universe is in key role in TGD even at elementary particle length
scales.\\ b) Quantum TGD describes Universe, which is critical at quantum
level. The breakthrough through the realization  that p-adic numbers
\cite{padrev} might provide technical tool for realizing criticality in
four dimensions. There is p-adic number field associated with each prime
and the algebraic extension allowing square for p-adic numbers
sufficiently near p-adic real axis is four-dimensional space and can be
regarded as p-adic counterpart of  future lightcone!   p-Adic conformal
invariance makes sense and this leads to the idea that p-Adic conformal
field theory should describe the quantum  field theory limit of TGD.  In
fact, there are many p-adic  field theories: a priori any prime can
correspond to effective long length scale topology for critical
system and various levels of topological condensate are assumed to
correspond to various values of $p$ in increasing hierarchy
$<p_1<p_2<..$.  \\ c) p-Adic numbers can be mapped in canonical manner
($\sum_n x_np^n \rightarrow \sum_nx_np^{-n}$) to real numbers and it was
found that p-adically analytic functions correspond typically to fractal
like structures.  From the fact that 2-adic level is fundamental it was
deduced that structures with sizes $2^nL_0$ (period doubling!)  should
appear in p-adic physics and from this that primes near powers $2$, in
particular prime powers of $2$, should be especially interesting
physically.  Mersenne primes $M_n=2^n-1$ are the primes nearest to powers
of $2$ and the hypothesis that fundamental elementary particle mass scales
correspond  to Mersenne primes was found to work \cite{padTGD}.

\vm

In this work p-adic description of Higgs mechanism is developed. The
 basic assumptions are following. \\
a) p-Adic Super Virasoro invariance fixes the general form of
 mass spectrum completely.  Combined with the idea that ordinary
 $H=M^4_+\times CP_2$ spinors should generalize to Kac Moody spinors this
leads to unique theory.  The special geometric features of $H$ allow to
construct Kac Moody spinors as tensor power of Ramond  (fermions) or
Neveu-Schwartz (bosons)
 $so(4)$ Super Virasoro representations \cite{Goddard,Bible}.  Color and
$u(1)$ degrees
 of freedom are described in terms of Super Virasoro representations and
the special properties of $su(3)$ are in decisive role also now.\\ b) The
fundamental description of Higgs mechanism is thermodynamical. Massivation
results from the thermal mixing of Planck mass excitations. Thermalization
takes place for Virasoro generator $L^0$ (proportional to mass squared).
Number theoretic constraints imply that p-adic temperature is quantized
($1/T=n$) in low temperature phase: allowed Boltzmann factors $exp(-kT)$
correspond to  integer powers $p^n$.    The  typical mass  squared scale
is predicted to be $M^2_0/p$, where $M^2_0$ is mass scale of order
$10^{-6}/G$ and together
 with Mersenne prime hypothesis leads to prediction of elementary particle
mass scales.\\
c) A more phenomenological description of Higgs mechanism is as
 breaking of super conformal symmetry: $(c,h)=(0,0)\rightarrow (c,h)\neq
(0,0)$
 and a tentative hypothesis is that the values of conformal symmetry
breaking parameters can be deduced from the thermodynamical approach.  \\
d) The hypothesis about genus-generation correspondence states that
elementary particle families correspond to  various boundary topologies
for the 3-surface associated particle correspendence turns out to be
decisive for understanding of elementary particle masses.  The boundary
contribution to mass squared is sum of cm contribution (boundary cm) and
modular contribution from conformally invariant degrees of freedom.  The
most pleasant surprise of long and painful calculation full of unpleasant
surprises was that
  a  general mass formula for modular contribution  follows by a few line
argument from the general form of  the elementary particle vacuum
functionals constructed in \cite{TGD}  to describe the conformally
invariant dynamics of modular  degrees of freedom.  The calculation of cm
contribution necessitates p-adic  thermodynamics for Kac Moody spinors
(analog of H-spinors) and is the hard part of the calculation. \\ e) The
general form of thermodynamic  mass formula is
 $M^2\propto  X(1)p+X(2)p^2$,  where $X(1)=\frac{D(1)}{D(0)}$ and $X(2)=
(2D(2)-\frac{D(1)^2}{D(0)})/D(0)$ are coefficients expressible in terms of
degeneracies $D(i)$ of $M^2=i=0,1,2$ states.  It turns out that $D(0)$, the
degeneracy of massless states,  is larger than one due to the fact that
ground state is tachyonic (note the analogy with Higgs mechanism) and for
temperature $T=1$ particle is light only provided  $D(1)/D(0)$ is integer
multiple of some
 constant.  $O(p^2)$ contribution to mass  is negligible if $X(2)$ is
integer multiple of
 same constant.

\vm

The practical  calculation of elementary particle  and hadron masses  not
an easy task since a lot of new mathematics
 is needed in precise practically working form.  What must be new to the
reader
 accustomed to live in real than p-adic world   is the fact p-adic
existence requirement (say for p-adic square root)  plays key role in
several deductions with highly nontrivial physical consequences. To
mention only temperature quantization at low temperature limit and
four-dimensionality of the extension allowing square root for p-adically
real number.  Also the nontrivialities associated with the correspondence
between real and p-adic numbers give extremely strong constraints. The
condition on the ratio $D(1)/D(0)$  as
 already mentioned. If the topological mixing matrices determining CKM
matrix are  rational  they   satisfy
 strong number theoretical   constraints.
 With suitable conditions on prime $p$ the square root of $M^2=-3$ (mass
of would be tachyon) is ordinary p-adic number rather than imaginary
p-adic number!  And so on....

\vm

In order to motivate the reader to go through the constructions and
calculations it is perhaps to state the main results:\\ a) Charged lepton
masses are predicted with accuracy better than one percent from the
contribution of the leptonic boundary  component.   Neutrino masses are
predicted apart from the uncertainty associated with their
 condensation levels and various bounds on neutrino masses leave only few
possible
 condensation levels  under considerations.  The existence of
leptohadrons \cite{Lepto,Heavy}, that
 is color bound states of color excited leptons, comes as a rigorous
prediction of TGD.\\ b)
 $W$ boson mass is predicted with accuracy better than one per cent.
Weinberg angle is predicted
 to be $3/8$, which introduces error of order $10$ per cent to $Z^0$
mass.  Taking into
 account the topological mixing for leptons and Coulomb corrections allows
to reproduce
 charged lepton and  gauge boson masses exactly.
 Photon and gluon are massless particles and hundreds of exotic bosons
turn out to have either Planck mass or to be light but massive (masses
are at least of order intermediate gauge boson mass) with exception of
 few colored exotic bosons. In TGD:eish Higgs
mechanism  there is no counterpart for the  Higgs doublet of standard
model. The operator having closest resemblance to Higgs field is Virasoro
generator $L_0$.\\ c)  Boundary
 contributions to quark masses can be calculated readily  (mass formula is
almost identical
 with the leptonic one!)  and are the dominating contributions to heavy
 quark masses.  Topological mixing of U and D quarks gives rise to CKM
matrix and  simple empirical inputs plus rationality assumption for the
mixing matrix elements   imply  number  theoretical congruences for the
parameters defining the mixing matrices. As
 a by-product a simple solution to the spin crisis of the proton
emerges.     It is possible
 to reproduce   physical CKM matrix and CP breaking is a number theoretic
necessity.  At this
 stage  it is not possible to predict contribution from   color Coulombic
interaction,   color
 hyper fine splittings, etc.   associated with the interior of the
hadronic surface.
 The values of the unknown parameters related to  these contributions  are
 however  deducible by comparison with hadron mass spectrum and by number
theoretic constraints  and mass spectrum can be reproduced with errors
below one per cent. Even  isospin-spin splittings
 can be reproduced  in terms of known interactions.

\vm

   There is one  notable exception. Top quark should have mass of order
$871 \ GeV$  if it really corresponds to $g=2$ quark.  If the observed top
candidate actually corresponds to $g=0$ $u$  quark of  $M_{89}$ hadron
physics ($M_{89}$  hadrons are tiny creatures condensed on ordinary quarks
and gluons!)
 as
 suggested already in \cite{padTGD} then the effective mass of the top is
obtained by scaling $m(u)=m(p)/3$ by factor $\sqrt{M_{107}/M_{89}}=512$:
the prediction is correct within 5 per cent accuracy!

\vm

The plan of the paper  series is following. \\
Part I.   It is shown that  the existence of p-adic
square root near  p-adic real axis leads to 4-dimensional algebraic
extension for $p>2$, which can be  regarded as the p-adic counter part of
future light cone. p-Adic light  cone consists of convergence cubes of square
root function and later work has shown that single convergence cube serves as
natural
region for the formulation of p-adic field theory limit of TGD.
\\ Part II:
 The general theory  of Higgs mechanism is developed. Thermal (the
fundamental)  and  conformal (the phenomenological) descriptions of Higgs
mechanism are   formulated.  The needed  Super Virasoro repsesentations
associated with Kac Moody spinors are constructed and  the concept of
massive Dirac operator  together with p-adic mechanism of tachyon
elimination is  introduced. General mass formulas for representations with
broken conformal symmetry are  derived.    The calculations for thermal
expectation values of mass squared operator are  described at general
level.\\
Part III.   The detailed calculations of lepton, quark and gauge boson
masses are carried
 out. Also the mass spectrum of exotic particles is derived. \\
Part IV.  The predictions for lepton and gauge boson are discussed in
detail and Coulomb corrections as well
 as topological mixing effects for charged lepton masses are discussed.
Hadronic
 mass formula is derived by modelling the yet unpredictable interior
contributions using simple
 empirical inputs plus number theory. Topological mixing and  CKM matrix
are discussed.   The reader
 interested mainly in  predictions  can perhaps  start the  reading from
this   paper.
  \\ Part V. TGD predicts color for leptons and quarks massless color
excitations as well as some exotic colored bosons. It is shown that
asymptotic freedom in standard sense is obtained.  A more quantitative
formulation for the
  TGD:eish condensate
hierachy with one p-adic  QCD associated with  each Mersenne prime
\cite{padTGD} is given.
 $Z^0$ width excludes light exotic fermions in real context but in p-adic
context the constraint is shown to fix the number of light fermions of
given
 type  only modulo certain integer.  The arguments necessitate a detailed
consideration of
 p-adic probability and unitarity concepts: the discussion  is left to
this paper although it might have had a more proper place in the first
part of the paper.
 Signatures of $M_{89}$ hadron Physics plus the hypothesis that the
observed top candidate is related to the decays of $\omega$ meson of
$M_{89}$ Physics are discussed.

	\section{p-Adic square root function and  p-adic light cone  }

The corner stone of the p-adic TGD is the following observation related to
the existence of square root of real p-adic number:  the algebraic
extensions of p-adic numbers allowing square root  near p-adic real axis
have dimension $D=4$ in $p>2$-adic case and dimension $D=8$ in
 2-adic case.  The proof of the result was based on the Taylor expansion
of square root and number theoretic facts.   Unfortunately the question
about the radius of the convergence was not considered at all.  The worst
situation would be that convergence radius is zero!
  This is not the case and  the region of spacetime allowing square root
function can be regarded as the p-adic counterpart of the future
 light cone. The result   demonstrates nice consistency of p-adic  TGD
with  the general assumptions of  TGD
 and  that big bang cosmology  might be related to  the existence of p-adic
square root!

\subsection{p-Adic square root function for $p>2$}

 The study of  the properties of the series representation of square
 root function shows that  the definition of square root  function is
possible in certain region around real p-adic axis. What is nice that this
region
 can be regarded as the {\bf p-adic counterpart of the future light cone
\/}  defined
 by the condition

\begin{eqnarray}
N_p(Im(Z))&<& N_p(t=Re(Z))=p^k
\end{eqnarray}

\noindent where the real p-adic  coordinate $t=Re(Z)$  is identified as
time  coordinate  and the imaginary    part of the p-adic coordinate is
identified as spatial coordinate.  p-Adic norm for four-dimensional
extension is analogous  to ordinary Euclidian distance.      p-Adic light
cone  consists of 'cylinders'  parallel  to time axis having radius
$N_p(t)= p^k$  and length $p^{k-1}(p-1)$: at points $t= p^k$.  As a real
space (recall the canonical correspondence)  the cross section of the
cylinder corresponds to parallelpiped  rather than ball.

\vm

The result can be understood heuristically as follows. \\ a) For
four-dimensional extension allowing square root  ($p>2$) one can
 construct square root at each p-adically real point $x(k,s)= sp^k$,
$s=1,...,p-1$, $k\in Z$.  The task is to show that  by using Taylor
expansion one can define square root also in some neighbourhood of each of
these points and find the form of this neighbourhood.  \\ b)  Using the
general series expansion of the  square root function one finds that the
convergence region is p-adic ball   defined by the condition

\begin{eqnarray}
N_p(Z-sp^k) &\leq &R(k)
\end{eqnarray}

\noindent and having    radius $R(k) = p^d, d \in Z$ around the expansion
point. \\
 c) A purely p-adic feature is that the {\bf  convergence spheres
associated with two points are either disjoint or identical\/}!   In
particular, the convergence sphere $B(y)$ associated with any  point
inside convergence sphere $B(x)$  is identical with $B(x)$:  $B(y)=
B(x)$.  The result follows directly  from the ultrametricity  of the
p-adic norm \cite{padTGD}.  The result means that stepwise analytic
continuation is not possible and one can construct square root function
only  in the {\bf union of   p-adic convergence spheres associated with
the p-adically real points\/} $x(k,s)=sp^k$. \\  d)  By the {\bf scaling
properties of the square root function\/} the convergence radius
$R(x(k,s))\equiv R(k)$ is related to $R(x(0,s))\equiv R(0)$  by the
scaling factor $p^{-k}$:

\begin{eqnarray}
R(k) &=& p^{-k}R(0)
\end{eqnarray}

\noindent so that {\bf convergence sphere expands  as a function of p-adic
time coordinate\/}.   The study of convergence reduces to the study of
the series at points $x=s=1,...,k-1$ with unit p-adic norm.  \\ e) {\bf
Two neighbouring
 points $x=s$ and $x=s+1$ cannot belong to  same convergence sphere\/}:
this would lead to contradiction with basic results of about square root
function at integer points.  Therefore the convergence radius  satisfies
the condition

\begin{eqnarray}
R(0)&<&1
\end{eqnarray}

\noindent The requirement that {\bf convergence is achieved at all points
 of the real axis\/} implies

\begin{eqnarray}
R(0)&=&\frac{1}{p}\nonumber\\
R(p^ks)&=& \frac{1}{p^{k+1}}
\end{eqnarray}

\noindent  If the convergence radius is indeed this then the region,  where
 square root is defined corresponds to a  connected light cone like region
defined by the condition $ N_p(Im(Z))= N_p(Re(Z))$ and
 $p>2$-adic space time is  p-adic counterpart of $M^4$ light cone.  If
 convergence radius is smaller the convergence region reduces to a union of
disjoint p-adic spheres with increasing radii.

\vm

 {\bf How the p-adic light cone differs from the  ordinary light cone \/}
 can be seen by studying the explicit form of the p-adic norm for $p>2$
square root allowing extension $Z=x+iy+\sqrt{p}(u+iv)$

\begin{eqnarray}
N_p(Z) &=& (N_p(det(Z)))^{\frac{1}{4}}\nonumber\\
&=& (N_p((x^2+y^2)^2+2p^2((xv-yu)^2+
(xu-yv)^2)+p^4(u^2+v^2)^2))^{\frac{1}{4}}\nonumber\\
\
\end{eqnarray}

\noindent where $det(Z)$ is the determinant of the linear map defined by
multiplication with $Z$.  The definition of convergence sphere for $x=s$
reduces to

\begin{eqnarray}
N_p(det(Z_3))&=&N_p(y^4+2p^2y^2(u^2+v^2)+p^4(u^2+v^2)^2))<1
\end{eqnarray}

\noindent  For physically interesting case $p \ mod \  4=3$ the points
$(y,u, v)$ satisfying the conditions

\begin{eqnarray}
 N_{p}(y)&\leq&\frac{1}{p}\nonumber\\ N_p(u)&\leq& 1
\nonumber\\ N_p(v)&\leq &1  \end{eqnarray}

\noindent belong to the sphere of convergence:  it is essential that
 for all $u$ and $v$ satisfying the conditions one has also
 $N_p(u^2+v^2)\leq 1$.   By the canonical correspondence  $\sum
x_np^n\rightarrow \sum x_np{-n}$
 between p-adic and real numbers \cite{padTGD} the real counterpart of the
sphere $r=t$ is now parallelpiped $0\leq y<1,0\leq u<p,0\leq v<p$,  which
expands with average velocity of light in discrete steps at times
$t=p^k$.

\vm

 The   emergence of  p-adic light cone as a  natural  p-adic coordinate
space
 is in nice  accordance with the basic assumptions about the imbedding
space of TGD and suggests  that {\bf big bang cosmology is related to the
existence of p-adic square root\/}!  The results gives  support for the
idea that p-adicity is responsible for the {\bf generation of lattice
structures\/} (convergence region for any function is expected to be more
or less parallelpiped like region).

\vm

A peculiar  feature of the  p-adic light cone is the instantaneous expansion
of 3-space at moments $t_p= p^k$.  A possible physical interpretation is
that {\bf p-adic light cone  or rather an individual convergence cube of the
 light
cone represent the time development of single maximal
quantum coherent region at p-adic level of topological condensate\/} (probably
there are many of them).  The instantaneous scaling of the size of region by
factor $\sqrt{p}$ at moment
 $t_R= p^{k/2}$ corresponds to a {\bf phase transition\/} and thus  to quantum
 jump.    Experience with p-adic QFT indeed shows that $L_p=\sqrt{p}L_0$
appears as
 infrared cutoff length for p-adic version of standard model.  The idea
that larger p-adic length
scales $p^{k/2}L_p$ would form quantum coherent regions for physically most
interesting values of
$p$ is perhaps unrealistic and $L_p$ probably gives a typical size of
 3-surface at p:th
condensate level.

\vm

   $p=M_{127}$,  the largest physically interesting Mersenne prime,
 provides an interesting example:\\
i) $p=M_{127}$-Adic light cone does not make sense for time and length scales
smaller than length scale defined by electron Compton length and QFT below this
length scale makes sense.  \\ ii) The
first phase transition would happen at time of order $10^{-1}$ seconds, which
corresponds to length scales of order $10^7 $ meters,  accidentally the
radius of a typical
planet.   \\ iii) The next phase transition takes place at
$t_R\simeq 10^{11}$
light years and is larger than  the present  age of the Universe.

\vm

Recent work with the p-adic field theory limit of TGD has shown that the
convergence cube of
p-adic
square root function having size $L_p= \frac{L_0}{\sqrt{p}}$,
$L_0\simeq  10^4\sqrt{G}$,   serves as
a natural quantization volume for p-adic counterpart of standard model.
  An open question is
whether also larger convergence cubes serve as quantization volumes or
whether $L_p$ gives
natural upper bound for the size of p-adic 3-surfaces. The original
idea \cite{padTGD} that
p-adic
manifolds, constructed by gluing together pieces of p-adic light cone together
 along their
sides, could be used to build Feynmann graphs with lines thickened to
4-manifolds  has
turned out to be not useful for physically most interesting (large) values
of $p$.

\subsection{Convergence radius for square root function}

In the following it will be shown that the convergence radius of
$\sqrt{t+Z}$
 is indeed nonvanishing for $p>2$.  The expression for the Taylor series
of   $\sqrt{t+Z}$ reads as

\begin{eqnarray}
\sqrt{t+Z}&=& = \sqrt{x}\sum_n a_n\nonumber\\
a_n&=& (-1)^n\frac{(2n-3)!!}{2^nn!} x^n\nonumber\\
x&=&\frac{Z}{t}
 \end{eqnarray}

\noindent The necessary criterion for the convergence is that the terms of
 the power series approach to zero at the limit $n\rightarrow \infty$.
The p-adic norm of $n$:th term is for $p>2$ given by

\begin{eqnarray}
N_p(a_n)&=& N_p(\frac{(2n-3)!!}{n!}) N_p(x^n)<N_p(x^n)N_p(\frac{1}{n!})
\end{eqnarray}

\noindent The dangerous term is clearly the $n!$ in the denominator.  In
the  following it will be shown that the condition

\begin{eqnarray}
U&\equiv &\frac{N_p(x^n)}{N_p(n!)} <1 \ for \ \ N_p(x)<1
\end{eqnarray}

\noindent holds true. The strategy is as follows:\\ a) The norm  of $x^n$
can be calculated trivially:
 $ N_p(x^n) =p^{-Kn}, K\ge 1$.\\ b)   $N_p(n!) $ is  calculated and an
upper bound for $U$ is derived at the limit of large $n$.

\subsubsection{p-Adic norm of $n!$ for $p>2$}

Lemma 1: Let $n= \sum_{i=0}^{k}n(i)p^i$,  $ 0\le n(i)<p$ be the p-adic
expansion of $n$. Then  $N_p(n!)$  can be expressed in the form

\begin{eqnarray}
N_p(n!)&=& \prod_{i=1}^{k} N(i)^{n(i)}\nonumber\\
N(1)&=&\frac{1}{p}\nonumber\\
N(i+1)&=& N(i)^{p-1}p^{-i}
\end{eqnarray}

\noindent  An explicit expression for $N(i)$ reads as

\begin{eqnarray}
N(i)&=& p^{-\sum_{m=0}^{i}  m(p-1)^{i-m}   }
\end{eqnarray}

\noindent Proof: $n!$ can be  written as a product

\begin{eqnarray}
N_p(n!) &=& \prod_{i=1}^{k}X(i,n(i) )\nonumber\\ X(k,n(k))&=&
N_p((n(k)p^k)!)\nonumber\\
X(k-1,n(k-1))&=&N_p(\prod_{i=1}^{n(k-1)p^{k-1}}(n(k)p^k+i))= N_p(
(n(k-1)p^{k-1})!)\nonumber\\
X(k-2,n(k-2))&=&N_p(\prod_{i=1}^{n(k-2)p^{k-2}}(n(k)p^k+n(k-1)p^{k-1}+i
))\nonumber\\ &=& N_p((n(k-2)p^{k-2})!)\nonumber\\ X(k-i,n(k-i))&=&
N_p((n(k-i)p^{k-i})!)
 \end{eqnarray}

\noindent  The factors $X(k,n(k))$ reduce in turn
to the form

\begin{eqnarray}
X(k,n(k))&=& \prod_{i=1}^{n(k)}Y(i,k )\nonumber\\
Y(i,k)&=& \prod_{m=1}^{p^k} N_p(ip^k+m)
\end{eqnarray}

\noindent The factors $Y(i,k)$ in turn are indentical and one has

\begin{eqnarray}
X(k,n(k))&=& X(k)^{n(k)}\nonumber\\
 X(k)&=& N_p(p^k!)
\end{eqnarray}

  The recursion formula for the factors $X(k)$ can be derived by  writing
explicitely the expression of $N_p(p^k!)$ for a few lowest values of $k$:\\
  1) $X(1)= N_p(p!) = p^{-1}$\\ 2) $ X(2) = N_p(p^2!)= X(1)^{p-1}p^{-2} $
( $p^2!$ decomposes to  $p-1$
 products having same norm as $p!$ plus the last term equal to $p^2$.\\ i)
$ X(i)= X(i-1)^{p-1}p^{-i}$

\vm

Using the recursion  formula repeatedly the explicit form of  $X(i)$ can
be derived easily. Combining the results one obtains for $N_p(n!)$ the
expression

\begin{eqnarray}
N_p(n!)&=& p^{-\sum_{i=0}^{k}n(i) A(i)}\nonumber\\
A(i)&=& \sum_{m=1}^{i} m(p-1)^{i-m}
\end{eqnarray}

\noindent  The sum $A(i)$  appearing in the exponent  as the coefficient of
 $n(i)$ can be calculated by using geometric series

\begin{eqnarray}
A(i)&=& (\frac{p-1}{p-2})^2
(p-1)^{i-1}(1+ \frac{i}{(p-1)^{i+1}}- \frac{(i+1)}{(p-1)^i})\nonumber\\
&\leq& (\frac{p-1}{p-2})^2(p-1)^{i-1}
\end{eqnarray}

\subsubsection{Upper bound for $N_p(\frac{x^n}{n!})$ for $p>2$}

By using the expressions $n= \sum_i n(i)p^i$, $N_p(x^n)=p^{-Kn}$ and the
expression of $N_p{n!}$ as well as the upper bound

\begin{eqnarray}
A(i)
&\leq& (\frac{p-1}{p-2})^2(p-1)^{i-1}
\end{eqnarray}

\noindent for $A(i)$ one obtains the  upper bound

\begin{eqnarray}
N_p(\frac{x^n}{n!}) \leq p^{-\sum_{i=0}^{k}n(i) p^i(K-(\frac{(p-1)}{(p-2)} )^2
(\frac{(p-1)}{p})^{i-1})} \nonumber\\
 \
\end{eqnarray}

\noindent It is clear that for $N_p(x)<1$  that is $K\ge1$ the upper
 bound goes to zero. For $p>3$ exponents are negative for all values of
$i$:  for $p=3$ some lowest exponents have wrong sign but this does not
spoil the convergence.  The convergence of the series is also obvious
since  the real valued series $\frac{1}{1-\sqrt{N_p(x)}}$ serves as
majorant.

\subsection{$p=2$ case}

In $p=2$ case the norm of a general term in the series of the square root
function can be calculated easily using the previous result for the norm of
$n!$:

\begin{eqnarray}
N_p(a_n)&=& N_p(\frac{(2n-3)!!}{2^nn!}) N_p(x^n)=  2^{-(K-1)n+\sum_{i=1}^{k}
n(i)\frac{i(i+1)}{2^{i+1} }}
\end{eqnarray}

\noindent At the limit $n\rightarrow \infty$  the sum term appearing in the
 exponent approaches zero and  convergence condition gives $ K>1$  so that
one has

\begin{eqnarray}
N_p(Z)&\equiv& (N_p(det(Z)))^{\frac{1}{8}}\leq\frac{1}{4}
\end{eqnarray}

\noindent  The result does  not imply disconnected set of convergence for
 square root function since the square root for half odd integers exists:

\begin{eqnarray}
\sqrt{s+\frac{1}{2}}= \frac{\sqrt{2s+1}}{\sqrt{2}}
\end{eqnarray}

\noindent  so that one can develop square as series in all half odd integer
 points of p-adic real axis.   As a consequence the structure for the set
of convergence is just the 8-dimensional counterpart of the p-adic light
cone.  {\bf Spacetime has natural binary structure\/} in the sense  that
each  $N_p(t)=2^k$ cylinder consists of two identical p-adic 8-balls
(parallelpipeds as real spaces).  Since $\sqrt{Z}$ appears in the
definition of the fermionic Ramond fields  one might wonder whether once
could  interpret  this binary structure as a geometric representation of
half odd integer spin.  The coordinate space associated with spacetime
representable  as a four-dimensional subset of this light cone inherits
the light cone structure.

\begin{center}
{\bf Acknowledgements\/}
\end{center}

\vm

It would not been possible to carry out this work without the  concrete
help of
 my friends  in concrete problems of the everyday life and I want to
express my gratitude to  them.  Also I want to thank J. Arponen,
 R. Kinnunen and J.  Maalampi
    for practical help and interesting discussions.

\end{document}